\documentclass[english]{extarticle}
\usepackage[T1]{fontenc}
\usepackage[latin9]{inputenc}
\usepackage{array}
\usepackage{units}
\usepackage{multirow}
\usepackage{amsmath}
\usepackage{amsthm}
\usepackage{setspace}

\makeatletter

\providecommand{\tabularnewline}{\\}

\newcommand{\lyxaddress}[1]{
\par {\raggedright #1
\vspace{1.4em}
\noindent\par}
}
\theoremstyle{plain}
\newtheorem{thm}{\protect\theoremname}[section]
  \theoremstyle{definition}
  \newtheorem{defn}[thm]{\protect\definitionname}

\usepackage[all]{xy}

\usepackage{cite}

\usepackage{placeins}

\usepackage{lineno}

\newcommand{\CHSH}{\mathsf{CHSH}}
\newcommand{\ICC}{\mathsf{ICC}}

\def\frontmatter@abstractheading{}

\date{}

\makeatother

\usepackage{babel}
  \providecommand{\definitionname}{Definition}
\providecommand{\theoremname}{Theorem}

\begin{document}

\title{On Contextuality in Behavioral Data}

\author{Ehtibar N. Dzhafarov\textsuperscript{1}, Janne V. Kujala\textsuperscript{2},
V\'ictor H. Cervantes\textsuperscript{1}, Ru Zhang\textsuperscript{1},
\\and Matt Jones\textsuperscript{3}}

\maketitle

\lyxaddress{\begin{center}
\textsuperscript{1}Purdue University, ehtibar@purdue.edu\\\textsuperscript{2}University
of Jyv\"askyl\"a, jvk@iki.fi\\\textsuperscript{3}University of
Colorado, mcj@colorado.edu
\par\end{center}}
\begin{abstract}
Dzhafarov, Zhang, and Kujala (Phil. Trans. Roy. Soc. A 374, 20150099)
reviewed several behavioral data sets imitating the formal design
of the quantum-mechanical contextuality experiments. The conclusion
was that none of these data sets exhibited contextuality if understood
in the generalized sense proposed in Dzhafarov, Kujala, and Larsson
(Found. Phys. 7, 762-782, 2015), while the traditional definition
of contextuality does not apply to these data because they violate
the condition of consistent connectedness (also known as marginal
selectivity, no-signaling condition, no-disturbance principle, etc.).
In this paper we clarify the relationship between (in)consistent connectedness
and (non)contextuality, as well as between the traditional and extended
definitions of (non)contextuality, using as an example the Clauser-Horn-Shimony-Holt
(CHSH) inequalities originally designed for detecting contextuality
in entangled particles.
\end{abstract}

\section{Introduction}

This paper is based on two talks given at the conference \emph{Quantum
Theory: From Foundations to Technologies} organized by Andrei Khrennikov
at the Linnaeus University in Växjö, Sweden. The content of these
talks has been to a large extent published elsewhere \cite{DKL2015,DKL2015LNCS,DzhRuKuj,KDL2015,KDconjecture},
and the present paper focuses on one specific aspect of these talks:
the relationship between \emph{(in)consistent connectedness} and \emph{(non)contextuality}.
This focus was prompted by a recent extensive exchange of personal
communications involving a few of our colleagues and related to a
new experiment announced by Aerts and Sozzo \cite{AertsSozzo2015}.
The issue in question is by no means new: it was in fact raised and
discussed in Ref.\ \cite{DK_Topics}, using as an example an experiment
by Aerts, Gabora, and Sozzo \cite{Aerts}. Later, this issue has become
a central one in the development of our approach to contextuality,
called Contextuality-by-Default (CbD) \cite{DK2014Scripta,DK_PLOS_2014,DK_qualified,DKL2015,DKL2015LNCS,DzhRuKuj,KDconjecture,KDL2015,KujDzholdnew}.
It has become clear from the discussion in question, however, that
there are still serious disagreements about this issue. The aim of
this paper is to offer a resolution for these disagreements and to
dispel possible conceptual confusions. 

Although prompted by a discussion of Ref. \cite{AertsSozzo2015},
this paper is not meant to be a critique of that or any particular
paper. We use the experiment presented in Ref. \cite{AertsSozzo2015}
and the paradigm in which it was conducted only as an example, one
providing an opportunity to demonstrate the workings of our theory
of contextuality and to make our points. We would like therefore to
play down the critical aspects of this paper.

\section{A list of important terms and notation conventions}

Special terms used in this paper are rigorously defined and every
notation convention is stipulated. The reader may, however, find it
useful to consult the following list from time to time to recall a
term or to more easily find where it is systematically discussed.

\paragraph*{Measurements (random variables).}

The generic notation for random variables is $R_{q}^{c}$, interpreted
as the measurement of \emph{property} $q$ in \emph{context} $c$.
If $c$ can be $c_{1},\ldots,c_{m}$ and $q$ can be $q_{1},\ldots,q_{n}$,
then instead of $R_{q_{i}}^{c_{j}}$ we write $R_{i}^{j}$ (see Section
\ref{sec:Systems-of-measurements}). 

In the ``Alice-Bob'' variant of cyclic-4 systems (Sections \ref{sec:Systems-of-measurements}
and \ref{sec:The-Alice-Bob-version}), we replace the general notation
$R_{q}^{c}$ for the measurements by $A$-$B$ notation, with the
following table of correspondences:

\begin{equation}
\begin{array}{cccccccc}
A_{11} & B_{11} & A_{12} & B_{12} & A_{21} & B_{21} & A_{22} & B_{22}\\
R_{1}^{1} & R_{2}^{1} & R_{1}^{4} & R_{4}^{4} & R_{3}^{2} & R_{2}^{2} & R_{3}^{3} & R_{4}^{3}
\end{array}.
\end{equation}
The logic of these correspondences is explained in Section \ref{sec:The-Alice-Bob-version},
(\ref{eq: A-B contexts}) and (\ref{eq: correspondences}).

\paragraph*{Bunches and connections.}

In a set of measurements: a subset of all $R_{q}^{c}$ with the same
$c$ and different $q$s forms a \emph{bunch of measurements} representing
context $c$; a subset of all $R_{q}^{c}$ with the same $q$ and
different $c$s forms a \emph{connection of measurements }representing
property $q$ (Section \ref{sec:Systems-of-measurements}).

\paragraph*{Consistent connectedness.}

Some connections have a certain property, (\ref{eq: intro consistency}),
that makes them \emph{consistent}; and a system of measurements with
all its connections consistent is \emph{consistently connected}. (See
Sections \ref{sec:Systems-of-measurements} and \ref{sec:Consistent-connectedness-and-contextuality}.)
The term is close to such terms as no-signaling, no-disturbance, etc.,
but void of their physical connotations.

\paragraph*{Contextual and noncontextual systems.}

\emph{(Non)contextuality} of a consistently connected system of measurements
is defined in Section \ref{sec:Traditional-understanding-of}, Definition
\ref{def consistently-connected}. The general definition of (non)contextuality,
for arbitrary systems, is given in Section \ref{sec:A-general-definition},
Definition \ref{def: inconsistent}.

\paragraph*{Couplings: $S$-notation and $T$-notation. }

The notion of (non)contextuality is based on the notion of a \emph{(probabilistic)
coupling}. Definition \ref{def: coupling for cyclic-4} defines the
couplings for cyclic-4 systems, and the subsequent remarks explain
how the definition applies to arbitrarily-connected systems of random
variables. 

When a coupling is constructed for \emph{all} measurements in a cyclic-4
system, then for each $R_{q}^{c}$ in this system we denote its counterpart
in the coupling by $S_{q}^{c}$ (Definition \ref{def: coupling for cyclic-4}). 

When a coupling is constructed for only a part of a cyclic-4 system,
specifically for pairs of the measurements $R_{q}^{c},R_{q}^{c'}$
forming a connection, then the corresponding elements of the coupling
are denoted $T_{q}^{c},T_{q}^{c'}$ (Definition \ref{def: couplings for connections}).

\section{\label{sec:Systems-of-measurements}Systems of measurements}

Of the two concepts characterizing a system of measurements, (in)consistent
connectedness and (non)contextuality, the former is about distributions
of the measurements of one and the same property in different contexts,
whereas the latter is about the (im)possibility of imposing certain
joint distributions on all the measurements, for all properties and
all contexts involved. 

Let a property $q$ be measured in contexts $c$ and $c'$. These
measurements can be denoted $R_{q}^{c}$ and $R_{q}^{c'}$. The property
$q$ may be a spin in a given particle along a given axis, and the
contexts $c$ and $c'$ may be defined by what other spins (say, in
other particles) are measured together with this one. Outside physics,
the property $q$ may be a question, and the contexts $c$ and $c'$
may be defined by whether this question was asked first or following
another question. Examples can be easily multiplied, within physics
and without. The measurements $R_{q}^{c}$ and $R_{q}^{c'}$ are two
different random variables, and their distributions can be the same
or different. If they are the same, we write 
\begin{equation}
R_{q}^{c}\sim R_{q}^{c'},\label{eq: intro consistency}
\end{equation}
and if this distributional equality holds for any $q$ and any contexts
$c,c'$ in which $q$ is measured, we say that the system of measurements
is \emph{consistently connected}. This term derives from the term
\emph{connection} that we use to denote a set of all measurements
of a given property in all contexts in which it is measured. For instance,
if property $q$ is measured in three contexts, $c,c',c''$, and in
no other contexts, then the set $\left\{ R_{q}^{c},R_{q}^{c'},R_{q}^{c''}\right\} $
is the connection for $q$. Consistent connectedness is known under
many different names: no-signaling condition \cite{PopescuRohrlich},
marginal selectivity \cite{DK_qualified,DK2012LNCS,TownSchw1989},
no-disturbance principle \cite{Ramanathan2012}, etc. (see Ref.~\cite{Cereceda2000}
for a few other terms). 

Contextuality is about all measurements $R_{q}^{c}$ composing a system.
Such a system can always be viewed as a set of \emph{bunches}, where
a bunch is defined as the set of all measurements made within a given
context. For instance, let $q,q',q''$ be measured in a context $c$
(e.g., $q,q',q''$ are three spins measured simultaneously, or three
questions asked of one and the same person), and let no other properties
be measured in that context. Then the set $\left\{ R_{q}^{c},R_{q'}^{c},R_{q''}^{c}\right\} $
is the bunch (of measurements) representing the context $c$. The
random variables within a bunch are jointly distributed, which means
that they can be viewed as a single (``vector-valued'') random variable. 

Assuming the numbers of the properties and the contexts are finite,
one can present the system of measurements in the form of a matrix,
in which rows correspond to the properties $\left\{ q_{1},\ldots,q_{n}\right\} $
and columns to the contexts $\left\{ c_{1},\ldots,c_{m}\right\} $,
and each cell $\left(i,j\right)$ is filled with the measurement $R_{i}^{j}$
if $q_{i}$ is measured in context $c_{j}$ (and is left empty otherwise).
\begin{equation}
\begin{array}{cccc}
 & \vdots\\
\cdots & R_{i}^{j} & \cdots & \textnormal{connection for }q_{i}\\
 & \vdots\\
 & \textnormal{bunch representing }c_{j}
\end{array}\label{eq:matrix}
\end{equation}
The random variables in any row of this matrix form a connection for
the corresponding property, and those in any column form a bunch representing
the corresponding context. 

We will focus in this paper on a special system of measurements, a
cyclic system of rank 4 \cite{DKL2015,KDL2015,KDconjecture}, or cyclic-4
system for short. Its best known implementation is the ``Alice-Bob''
version of the Einstein-Podolsky-Rosen-Bohm system (EPR-B, where B
can also stand for Bell). This system has been prominently studied
in relation to contextuality since John Bell's pioneering work \cite{Bell1964,Bell1966},
although the conceptual framework used in quantum physics (entanglement,
nonlocality) initially did not include contextuality explicitly. Outside
quantum physics the term ``nonlocality'' rarely makes sense, and
``entanglement'', if it does make sense (even if only metaphorical),
can always be taken as a possible ``explanation'' for contextuality,
if observed. We will return to the EPR-B implementation of the cyclic-4
system in Section \ref{sec:The-Alice-Bob-version}.

The system in question can be presented in the format of the matrix
(\ref{eq:matrix}) as follows: 
\begin{equation}
\begin{array}{lcccc}
 & c_{1} & c_{2} & c_{3} & c_{4}\\
q_{1} & R_{1}^{1} & \cdot & \cdot & R_{1}^{4}\\
q_{2} & R_{2}^{1} & R_{2}^{2} & \cdot & \cdot\\
q_{3} & \cdot & R_{3}^{2} & R_{3}^{3} & \cdot\\
q_{4} & \cdot & \cdot & R_{4}^{3} & R_{4}^{4}
\end{array}\label{eq: matrix EPRB}
\end{equation}
We will change the notation later, in Section \ref{sec:The-Alice-Bob-version},
to conform with the traditional interpretation of the properties and
contexts involved, e.g., in a pair of two entangled particles. For
now, one can think of any four properties measured in any four contexts
so that (a) each context contains two properties measured together;
(b) each property is measured in two different contexts; (c) no two
contexts share more than one property; and (d) each measurement is
a binary random variable, with values $\pm1$.

\section{Traditional understanding of contextuality in cyclic-4 systems}

The traditional understanding of contextuality in the cyclic-4 paradigm
can be presented as follows. Let us assume the measurements $R_{q}^{c},R_{q}^{c'}$
of any property $q$ in the contexts $c,c'$ in which it is measured
to be in fact one and the same random variable, $R_{q}^{*}$. This
assumption can be referred to as that of \emph{context-irrelevance},
and in many traditional treatments it is made implicitly, by the virtue
of indexing the measurements by the properties being measured but
not by the contexts. The assumption implies that our matrix (\ref{eq: matrix EPRB})
can be written as
\begin{equation}
\begin{array}{lcccc}
 & c_{1} & c_{2} & c_{3} & c_{4}\\
q_{1} & R_{1}^{*} & \cdot & \cdot & R_{1}^{*}\\
q_{2} & R_{2}^{*} & R_{2}^{*} & \cdot & \cdot\\
q_{3} & \cdot & R_{3}^{*} & R_{3}^{*} & \cdot\\
q_{4} & \cdot & \cdot & R_{4}^{*} & R_{4}^{*}
\end{array}\label{eq: matrix traditional}
\end{equation}
immediately and trivially implying consistent connectedness: e.g.,
$R_{1}^{*}$ measured together with $R_{2}^{*}$ in context $c_{1}$
is precisely the same random variable as $R_{1}^{*}$ measured together
with $R_{4}^{*}$ in context $c_{4}$ (otherwise they could not be
denoted by the same symbol $R_{1}^{*}$); and, of course, a fixed
random variable has a fixed distribution.

It is easy to show that if random variables are understood within
the framework of the classical, Kolmogorovian probability theory (KPT),
then the four random variables $\left\{ R_{1}^{*},R_{2}^{*},R_{3}^{*},R_{4}^{*}\right\} $
in system (\ref{eq: matrix traditional}) possess a joint distribution.
Indeed, the random variables $\left(R_{1}^{*},R_{2}^{*}\right)$ in
context $c_{1}$ are jointly distributed, which means that they are
two measurable functions defined on the same probability space $S$,
\begin{equation}
R_{1}^{*}:S\rightarrow\left\{ -1,+1\right\} ,\;R_{2}^{*}:S\rightarrow\left\{ -1,+1\right\} .
\end{equation}
{[}More precisely, $S$ is a set in a probability space $\left(S,\Sigma,\mu\right)$,
where $\Sigma$ is a sigma-algebra (set of events) on $S$ and $\mu$
some probability measure. A function $X:S\rightarrow\left\{ -1,+1\right\} $
is measurable (and therefore $X$ is a random variable) if the set
of values mapped into $+1$ is an event (i.e., it belongs to $\Sigma,$
and therefore has a well-defined probability value). We conveniently
confuse the set $S$ with the probability space containing $S$.{]} 

The random variables $\left(R_{1}^{*},R_{4}^{*}\right)$ in context
$c_{4}$ are also jointly distributed, whence $R_{1}^{*},R_{4}^{*}$
are measurable functions on the same probability space. This must
be the same space $S$ as above because the variable $R_{1}^{*}$
in the contexts $c_{1}$ and $c_{4}$ is the same. Hence
\begin{equation}
R_{4}^{*}:S\rightarrow\left\{ -1,+1\right\} .
\end{equation}
Finally, in context $c_{2}$, the random variables $\left(R_{2}^{*},R_{3}^{*}\right)$
are jointly distributed, and we conclude that
\begin{equation}
R_{3}^{*}:S\rightarrow\left\{ -1,+1\right\} .
\end{equation}
As a result, all four random variables in (\ref{eq: matrix EPRB})
are measurable functions defined on the same probability space, i.e.,
they are jointly distributed. 

{[}There is a naive way of arriving at the same conclusion, by assuming
that in the KPT any set of random variables is jointly distributed.
This view is untenable \cite{DK2014Scripta,DK_qualified}.{]}

Now, the joint distribution of $\left\{ R_{1}^{*},R_{2}^{*},R_{3}^{*},R_{4}^{*}\right\} $
is unobservable, because no two measurements made in two different
contexts (such as $\left\{ R_{1}^{*},R_{3}^{*}\right\} $ or $\left\{ R_{2}^{*},R_{4}^{*}\right\} $),
``co-occur'' in any empirical meaning of ``co-occurrence''. One
can only observe (i.e., estimate from observed frequencies of co-occurrences)
the distributions of four specific subsets of $\left\{ R_{1}^{*},R_{2}^{*},R_{3}^{*},R_{4}^{*}\right\} $,
the pairs of random variables forming the columns of matrix (\ref{eq: matrix traditional}).
We have the following theorem about these pairs that was first proved,
mutatis mutandis, in Ref.~\cite{9CHSH}. In its formulation, $\left\langle \cdot\right\rangle $
denotes expected value, and the maximum is taken over all combinations
of $+$ and $-$ signs replacing $\pm$ so that the number of the
$-$ signs is odd (1 or 3).
\begin{thm}
\label{thm: traditional}In any system described by (\ref{eq: matrix traditional}),
\begin{equation}
\max_{\begin{array}{c}
odd\:number\\
of-\textnormal{'s}
\end{array}}\left(\pm\left\langle R_{1}^{*}R_{2}^{*}\right\rangle \pm\left\langle R_{2}^{*}R_{3}^{*}\right\rangle \pm\left\langle R_{3}^{*}R_{4}^{*}\right\rangle \pm\left\langle R_{4}^{*}R_{1}^{*}\right\rangle \right)\leq2.\label{eq: CHSH stupid}
\end{equation}

\end{thm}
The inequality (\ref{eq: CHSH stupid}) is usually presented as a
necessary condition for the existence of a joint distribution of $\left\{ R_{1}^{*},R_{2}^{*},R_{3}^{*},R_{4}^{*}\right\} $,
implying that (\ref{eq: CHSH stupid}) can be violated, in which case
$\left\{ R_{1}^{*},R_{2}^{*},R_{3}^{*},R_{4}^{*}\right\} $ do not
have a joint distribution and we say that the cyclic-4 system is \emph{contextual.}
This understanding, however, lacks logical rigor. If the left-hand
side of (\ref{eq: CHSH stupid}) can be computed at all, then the
expected products $\left\langle R_{1}^{*}R_{2}^{*}\right\rangle ,\ldots,\left\langle R_{4}^{*}R_{1}^{*}\right\rangle $
in it are well-defined, whence each of the corresponding pairs $\left(R_{1}^{*},R_{2}^{*}\right),\ldots,\left(R_{4}^{*},R_{1}^{*}\right)$
has a well-defined joint distribution. But then, as we have seen,
the entire set $\left\{ R_{1}^{*},R_{2}^{*},R_{3}^{*},R_{4}^{*}\right\} $
has to have a joint distribution too, and then, by Theorem \ref{thm: traditional},
(\ref{eq: CHSH stupid}) must hold. It simply cannot be violated.

Put differently but equivalently, if $\left\{ R_{1}^{*},R_{2}^{*},R_{3}^{*},R_{4}^{*}\right\} $
do not possess a joint distribution, then at least two of the four
pairs $\left(R_{1}^{*},R_{2}^{*}\right),\ldots,\left(R_{4}^{*},R_{1}^{*}\right)$
forming columns of matrix (\ref{eq: matrix traditional}) do not have
joint distributions (because a global joint distribution follows from
any three of these pairs being jointly distributed). But if this is
the case, the left-hand side of (\ref{eq: CHSH stupid}) simply cannot
be computed.

\section{\label{sec:Consistent-connectedness-and-contextuality}Consistent
connectedness and contextuality in traditional understanding}

\begin{table}
\caption{\label{tab: Example 1}Example in which the left-hand side of (\ref{eq: CHSH stupid})
is seemingly well-defined and exceeds 2. Since this is mathematically
impossible, there should be a hidden assumption here that is false.
\medskip{}
}

\begin{footnotesize}

\begin{centering}
\begin{tabular}{cc}
 & \tabularnewline
 & \tabularnewline
\multirow{2}{*}{$R_{1}^{*}$} & \tabularnewline
 & \tabularnewline
 & \tabularnewline
\end{tabular}%
\begin{tabular}{c|c|c|c}
\multicolumn{1}{c}{} & \multicolumn{2}{c}{$R_{2}^{*}$} & \tabularnewline
\multicolumn{1}{c}{} & \multicolumn{1}{c}{} & \multicolumn{1}{c}{} & \tabularnewline
\cline{2-3} 
$ $ & $+1$ & $-1$ & \tabularnewline
\hline 
\multicolumn{1}{|c|}{$+1$} & $\nicefrac{1}{2}$ & 0 & \multicolumn{1}{c|}{$\nicefrac{1}{2}$}\tabularnewline
\hline 
\multicolumn{1}{|c|}{$-1$} & 0 & $\nicefrac{1}{2}$ & \multicolumn{1}{c|}{$\nicefrac{1}{2}$}\tabularnewline
\hline 
 & $\nicefrac{1}{2}$ & $\nicefrac{1}{2}$ & \tabularnewline
\cline{2-3} 
\end{tabular}$\qquad$$\qquad$%
\begin{tabular}{c|c|c|c}
\multicolumn{1}{c}{} & \multicolumn{2}{c}{$R_{4}^{*}$} & \tabularnewline
\multicolumn{1}{c}{} & \multicolumn{1}{c}{} & \multicolumn{1}{c}{} & \tabularnewline
\cline{2-3} 
 & $+1$ & $-1$ & \tabularnewline
\hline 
\multicolumn{1}{|c|}{$+1$} & $\nicefrac{1}{2}$ & 0 & \multicolumn{1}{c|}{$\nicefrac{1}{2}$}\tabularnewline
\hline 
\multicolumn{1}{|c|}{$-1$} & 0 & $\nicefrac{1}{2}$ & \multicolumn{1}{c|}{$\nicefrac{1}{2}$}\tabularnewline
\hline 
 & $\nicefrac{1}{2}$ & $\nicefrac{1}{2}$ & \tabularnewline
\cline{2-3} 
\end{tabular}
\par\end{centering}

\begin{centering}
\begin{tabular}{cc}
 & \tabularnewline
\multirow{2}{*}{$R_{3}^{*}$} & \tabularnewline
 & \tabularnewline
 & \tabularnewline
\end{tabular}%
\begin{tabular}{|c|c|c|c|}
\cline{2-3} 
\multicolumn{1}{c|}{} & $+1$ & $-1$ & \multicolumn{1}{c}{}\tabularnewline
\hline 
$+1$ & $\nicefrac{1}{2}$ & 0 & $\nicefrac{1}{2}$\tabularnewline
\hline 
$-1$ & 0 & $\nicefrac{1}{2}$ & $\nicefrac{1}{2}$\tabularnewline
\hline 
\multicolumn{1}{c|}{} & $\nicefrac{1}{2}$ & $\nicefrac{1}{2}$ & \multicolumn{1}{c}{}\tabularnewline
\cline{2-3} 
\end{tabular}$\qquad$$\qquad$%
\begin{tabular}{|c|c|c|c|}
\cline{2-3} 
\multicolumn{1}{c|}{} & $+1$ & $-1$ & \multicolumn{1}{c}{}\tabularnewline
\hline 
$+1$ & 0 & $\nicefrac{1}{2}$ & $\nicefrac{1}{2}$\tabularnewline
\hline 
$-1$ & $\nicefrac{1}{2}$ & 0 & $\nicefrac{1}{2}$\tabularnewline
\hline 
\multicolumn{1}{c|}{} & $\nicefrac{1}{2}$ & $\nicefrac{1}{2}$ & \multicolumn{1}{c}{}\tabularnewline
\cline{2-3} 
\end{tabular}
\par\end{centering}

\end{footnotesize}
\end{table}

One can be easily confused by the reasoning above, because it may
seem that it is trivial to construct a system (\ref{eq: matrix traditional})
in which all four expected products $\left\langle R_{1}^{*}R_{2}^{*}\right\rangle ,\ldots,\left\langle R_{4}^{*}R_{1}^{*}\right\rangle $
are well-defined while (\ref{eq: CHSH stupid}) is violated (and it
is routinely claimed that quantum mechanics predicts such situations
and experiments confirm these predictions). This seemingly trivial
possibility, however, is merely an illusion, because such a construction
would be one of a mathematically self-contradictory system. One example
is given by the four distributions in Table \ref{tab: Example 1},
where entries within the $2\times2$ interiors are joint probabilities,
while the margins show marginal probabilities. The expected products
here are
\[
\left\langle R_{1}^{*}R_{2}^{*}\right\rangle =\left\langle R_{2}^{*}R_{3}^{*}\right\rangle =-\left\langle R_{3}^{*}R_{4}^{*}\right\rangle =\left\langle R_{4}^{*}R_{1}^{*}\right\rangle =1,
\]
and the left-hand side of (\ref{eq: CHSH stupid}) is 4, violating
the inequality. As this system is mathematically impossible, we must
have made an assumption that this contradiction demonstrates to be
false.

What might this assumption be? Can it be that $R_{1}^{*}$ are $R_{2}^{*}$
are not jointly distributed, or that they are not well-defined random
variables? The answer is clearly negative: $R_{1}^{*}$ are $R_{2}^{*}$
are observed empirically and jointly. The same reasoning applies to
the remaining three pairs: in each pair the two random variables are
well-defined and jointly distributed. The only possible error therefore
is in the identity of these random variables across different contexts:
we have assumed, e.g., that $R_{3}^{*}$ measured together with $R_{2}^{*}$
is the same random variable as $R_{3}^{*}$ measured together with
$R_{4}^{*}$. It must be wrong to label the measurements by the measured
properties only, ignoring the contexts. 

This means that a correct initial representation of the system would
be as in Table \ref{tab: Example 1corr}, with the random variables
contextually labeled, so that the pairs of measurements forming different
bunches do not overlap. If one makes the assumption that, for any
property $q$ and any two contexts $c,c'$, the measurements $R_{q}^{c}$
and $R_{q}^{c'}$ in this matrix are ``one and the same variable''
$R_{q}^{*}$, then this assumption is rejected by reductio ad absurdum:
if it were correct, (\ref{eq: CHSH stupid}) would have to hold, and
it does not. 

\begin{table}
\caption{\label{tab: Example 1corr}Example of Table \ref{tab: Example 1}
without the assumption of context-irrelevance: the identity of measurements
(random variables) depends not only on what is measured but also on
the context in which it is measured.\medskip{}
}

\begin{footnotesize}

\begin{centering}
\begin{tabular}{ccc|c|c|c}
 &  & \multicolumn{1}{c}{} & \multicolumn{2}{c}{$R_{2}^{1}$} & \tabularnewline
 &  & \multicolumn{1}{c}{} & \multicolumn{1}{c}{} & \multicolumn{1}{c}{} & \tabularnewline
\cline{4-5} 
 &  &  & $+1$ & $-1$ & \tabularnewline
\cline{3-6} 
\multirow{2}{*}{$R_{1}^{1}$} & \multicolumn{1}{c|}{\multirow{2}{*}{}} & $+1$ & $\nicefrac{1}{2}$ & 0 & \multicolumn{1}{c|}{$\nicefrac{1}{2}$}\tabularnewline
\cline{3-6} 
 &  & $-1$ & 0 & $\nicefrac{1}{2}$ & \multicolumn{1}{c|}{$\nicefrac{1}{2}$}\tabularnewline
\cline{3-6} 
 &  &  & $\nicefrac{1}{2}$ & $\nicefrac{1}{2}$ & \tabularnewline
\cline{4-5} 
\end{tabular}$\qquad$$\qquad$%
\begin{tabular}{c|c|c|ccc}
\multicolumn{1}{c}{} & \multicolumn{2}{c}{$R_{4}^{4}$} &  &  & \tabularnewline
\multicolumn{1}{c}{} & \multicolumn{1}{c}{} & \multicolumn{1}{c}{} &  &  & \tabularnewline
\cline{2-3} 
 & $+1$ & $-1$ &  &  & \tabularnewline
\cline{1-4} 
\multicolumn{1}{|c|}{$+1$} & $\nicefrac{1}{2}$ & 0 & \multicolumn{1}{c|}{$\nicefrac{1}{2}$} &  & \multirow{2}{*}{$R_{1}^{4}$}\tabularnewline
\cline{1-4} 
\multicolumn{1}{|c|}{$-1$} & 0 & $\nicefrac{1}{2}$ & \multicolumn{1}{c|}{$\nicefrac{1}{2}$} &  & \tabularnewline
\cline{1-4} 
 & $\nicefrac{1}{2}$ & $\nicefrac{1}{2}$ &  &  & \tabularnewline
\cline{2-3} 
\end{tabular}
\par\end{centering}

\begin{centering}
\begin{tabular}{ccc|c|c|c}
\cline{4-5} 
 &  &  & $+1$ & $-1$ & \tabularnewline
\cline{3-6} 
\multirow{2}{*}{$R_{3}^{2}$} & \multicolumn{1}{c|}{} & $+1$ & $\nicefrac{1}{2}$ & 0 & \multicolumn{1}{c|}{$\nicefrac{1}{2}$}\tabularnewline
\cline{3-6} 
 & \multicolumn{1}{c|}{} & $-1$ & 0 & $\nicefrac{1}{2}$ & \multicolumn{1}{c|}{$\nicefrac{1}{2}$}\tabularnewline
\cline{3-6} 
 &  &  & $\nicefrac{1}{2}$ & $\nicefrac{1}{2}$ & \tabularnewline
\cline{4-5} 
 &  & \multicolumn{1}{c}{} & \multicolumn{2}{c}{} & \tabularnewline
 &  & \multicolumn{1}{c}{} & \multicolumn{2}{c}{$R_{2}^{2}$} & \tabularnewline
\end{tabular}$\qquad$$\qquad$%
\begin{tabular}{c|c|c|ccc}
\cline{2-3} 
 & $+1$ & $-1$ &  &  & \tabularnewline
\cline{1-4} 
\multicolumn{1}{|c|}{$+1$} & 0 & $\nicefrac{1}{2}$ & \multicolumn{1}{c|}{$\nicefrac{1}{2}$} &  & \multirow{2}{*}{$R_{3}^{3}$}\tabularnewline
\cline{1-4} 
\multicolumn{1}{|c|}{$-1$} & $\nicefrac{1}{2}$ & 0 & \multicolumn{1}{c|}{$\nicefrac{1}{2}$} &  & \tabularnewline
\cline{1-4} 
 & $\nicefrac{1}{2}$ & $\nicefrac{1}{2}$ &  &  & \tabularnewline
\cline{2-3} 
\multicolumn{1}{c}{} & \multicolumn{2}{c}{} &  &  & \tabularnewline
\multicolumn{1}{c}{} & \multicolumn{2}{c}{$R_{4}^{3}$} &  &  & \tabularnewline
\end{tabular}
\par\end{centering}

\end{footnotesize}
\end{table}

Below we will present a rigorous way of formulating the hypothesis
that random variables measuring the same property in different contexts
are (in some sense) ``the same''. We already have, however, sufficient
clarity about this hypothesis to address the often misunderstood question
of the relationship, within the framework of this hypothesis, between
the concepts of consistent connectedness and contextuality. 

It is clear that the assumption of consistent connectedness can be
formulated and, in special cases, even justified without assuming
context-irrelevance. Its formulation for the cyclic-4 system presented
in the form (\ref{eq: matrix EPRB}) is
\begin{equation}
\begin{array}{cc}
R_{1}^{4}\sim R_{1}^{1}, & R_{2}^{1}\sim R_{2}^{2},\\
R_{3}^{2}\sim R_{3}^{3}, & R_{4}^{3}\sim R_{4}^{4}.
\end{array}\label{eq: cc}
\end{equation}
Such a hypothesis can often be entertained without assuming that the
identically distributed random variables are ``the same''. For instance,
in the classical entanglement paradigm for two electrons, property
1 corresponds to Alice's choice of a certain axis in her particle,
and the context $c_{1}$ is defined by Bob's simultaneously choosing
axis $2$ in his particle, while the context $c_{4}$ is defined by
Bob's simultaneously choosing axis 4 (on labeling Alice's two axes
1,3, and Bob's two axes 2,4). If the two particles are space-like
separated, one should assume that Bob's settings cannot influence
Alice's measurements, which implies the distribution of $R_{1}^{1}$
is the same as the distribution of $R_{1}^{4}$. No physical principle
prevents one, however, from viewing $R_{1}^{1}$ and $R_{1}^{4}$
as different random variables with one and the same distribution.
We have seen already that one's denying this view leads to a mathematical
contradiction. 

The expected products in (\ref{eq: CHSH stupid}) also can be written
without regard to the context-irrelevance hypothesis. One can replace
$\left\langle R_{1}^{*}R_{2}^{*}\right\rangle $ with $\left\langle R_{1}^{1}R_{2}^{1}\right\rangle $,
$\left\langle R_{2}^{*}R_{3}^{*}\right\rangle $ with $\left\langle R_{2}^{2}R_{3}^{2}\right\rangle $,
etc. to obtain the following analogue of inequality (\ref{eq: CHSH stupid}):
\begin{equation}
\max_{\begin{array}{c}
odd\:number\\
of-\textnormal{'s}
\end{array}}\left(\pm\left\langle R_{1}^{1}R_{2}^{1}\right\rangle \pm\left\langle R_{2}^{2}R_{3}^{2}\right\rangle \pm\left\langle R_{3}^{3}R_{4}^{3}\right\rangle \pm\left\langle R_{4}^{4}R_{1}^{4}\right\rangle \right)\leq2.\label{eq: CHSH}
\end{equation}
Using this formulation, Theorem \ref{thm: traditional} can be understood
as stating that (\ref{eq: CHSH}) holds under the context-irrelevance
hypothesis. If this hypothesis does not hold, (\ref{eq: CHSH}) does
not have to be satisfied and therefore cannot be derived as a theorem.
One can always check whether it holds or not, but the outcome has
no known to us interpretation if $R_{q}^{c}$ and $R_{q}^{c'}$ are
not assumed always to be the same. Now, the context-irrelevance hypothesis
simply cannot be entertained if consistent connectedness is violated:
``one and the same'' random variable cannot have two different distributions.
There is therefore no reason for checking the inequality (\ref{eq: CHSH})
if (\ref{eq: cc}) does not hold.

The only exception can be made in an imaginary situation wherein the
consistency of connectedness is not known (e.g., it is not established
in a statistically reliable way), but one knows (in a statistically
reliable way) that the inequality (\ref{eq: CHSH}) is violated. In
this case one can reject the context-irrelevance hypothesis by the
following reasoning: (i) if the system is consistently connected,
then the hypothesis of context-irrelevance leads to (\ref{eq: CHSH}),
which is rejected; (ii) if the system is not consistently connected,
the hypothesis of context-irrelevance is rejected as well; (iii) hence
this hypothesis is rejected.

\section{\label{sec:Traditional-understanding-of}Traditional understanding
of contextuality translated into the CbD language}

In accordance with the CbD approach, $R_{q}^{c}$ and $R_{q}^{c'}$
($c\not=c'$) are a priori different random variables, and since they
are never observed ``together'' (in any empirically grounded sense
of this word), they do not posses a joint distribution. The conceptual
coherence and advantages offered by this understanding of random variables
recorded in different contexts has been discussed in Refs.\ \cite{conversations,DK2014Scripta,DKL2015}.
In the framework of KPT this means that $R_{q}^{c}$ and $R_{q}^{c'}$
are functions defined on two different probability spaces:
\begin{equation}
R_{q}^{c}:S_{c}\rightarrow\left\{ -1,+1\right\} ,\quad R_{q}^{c'}:S_{c'}\rightarrow\left\{ -1,+1\right\} .
\end{equation}
It is therefore impossible to hypothesize that $R_{q}^{c}$ and $R_{q}^{c'}$
($c\not=c'$) are in fact ``the same''. Nor is it possible to treat
these $R_{q}^{c}$ and $R_{q}^{c'}$ as ``different but always equal
to each other'',
\begin{equation}
\Pr\left[R_{q}^{c}=R_{q}^{c'}\right]=1,\label{eq: equality stupid}
\end{equation}
since this statement also implies a joint distribution of $\left(R_{q}^{c},R_{q}^{c'}\right)$,
translating into $S_{c}=S_{c'}$. 

To formulate the analogue of the context-irrelevance hypothesis within
the framework of CbD one has to use the foundational notion of a \emph{(probabilistic)
coupling}. 
\begin{defn}
\label{def: coupling for cyclic-4}A coupling for the cyclic-4 system
(\ref{eq: matrix EPRB}) is a set of eight jointly distributed random
variables
\begin{equation}
\left(S_{1}^{1},S_{2}^{1},S_{2}^{2},S_{3}^{2},S_{3}^{3},S_{4}^{3},S_{4}^{4},S_{1}^{4}\right)\label{eq: coupling for EPR-B}
\end{equation}
such that
\begin{equation}
\begin{array}{cc}
\left(S_{1}^{1},S_{2}^{1}\right)\sim\left(R_{1}^{1},R_{2}^{1}\right), & \left(S_{2}^{2},S_{3}^{2}\right)\sim\left(R_{2}^{2},R_{3}^{2}\right),\\
\left(S_{3}^{3},S_{4}^{3}\right)\sim\left(R_{3}^{3},R_{4}^{3}\right), & \left(S_{4}^{4},S_{1}^{4}\right)\sim\left(R_{4}^{4},R_{1}^{4}\right).
\end{array}
\end{equation}

\end{defn}
In other words, the bunches of the system are distributed as the corresponding
marginals of the coupling. A system has generally an infinity of couplings.

The notion of a coupling is not confined to cyclic-4 systems. It applies
to any system of random variables, the idea being that (a) the coupling
is a set of jointly distributed random variables in a one-to-one correspondence
with the variables constituting the system being coupled; and (b)
the observable parts of this system are distributed in the same way
as the corresponding marginals (subsets, or \emph{subcouplings}) of
the coupling. In particular, the system being coupled can be a connection
of the cyclic-4 system.

Recall that the connection for property $q$ is the set of all random
variables measuring $q$ in different contexts. In the cyclic-4 system
the connection for property 1 is $\left\{ R_{1}^{1},R_{1}^{4}\right\} $,
for property $2$ it is $\left\{ R_{2}^{1},R_{2}^{2}\right\} $, etc.,
along the rows of the matrix (\ref{eq: matrix EPRB}). Each of these
connections taken in isolation has its couplings. 
\begin{defn}
\label{def: couplings for connections}A pair of jointly distributed
random variables $\left\{ T_{q}^{c},T_{q}^{c'}\right\} $ is a coupling
of a connection $\left\{ R_{q}^{c},R_{q}^{c'}\right\} $ in a cyclic-4
system if 
\begin{equation}
T_{q}^{c}\sim R_{q}^{c},\quad T_{q}^{c'}\sim R_{q}^{c'}.\label{eq: coupling for connection}
\end{equation}
The coupling $\left\{ T_{q}^{c},T_{q}^{c'}\right\} $ is called \emph{maximal}
if the probability with which $T_{q}^{c},T_{q}^{c'}$ attain equal
values, $\Pr\left[T_{q}^{c}=T_{q}^{c'}\right]$, is maximal among
all couplings of $\left\{ R_{q}^{c},R_{q}^{c'}\right\} $.
\end{defn}
Another way of stating the second part of the definition is that $\Pr\left[T_{q}^{c}=T_{q}^{c'}\right]$
is as large as it is allowed to be by the distributions of $T_{q}^{c}$
and $T_{q}^{c'}$, which are fixed by (\ref{eq: coupling for connection}).
The following theorem says that this concept is well-defined.
\begin{thm}[Refs.\ \cite{DKL2015,KDL2015}]
\label{thm: A-maximal-coupling}A maximal coupling $\left\{ T_{q}^{c},T_{q}^{c'}\right\} $
of a connection $\left\{ R_{q}^{c},R_{q}^{c'}\right\} $ in a cyclic-4
system exists and its distribution is unique: it is defined by (\ref{eq: coupling for connection})
and 
\begin{equation}
\left\langle T_{q}^{c}T_{q}^{c'}\right\rangle =1-\left|\left\langle T_{q}^{c}\right\rangle -\left\langle T_{q}^{c'}\right\rangle \right|=1-\left|\left\langle R_{q}^{c}\right\rangle -\left\langle R_{q}^{c'}\right\rangle \right|,
\end{equation}
or equivalently,
\begin{equation}
\Pr\left[T_{q}^{c}=T_{q}^{c'}\right]=1-\left|\Pr\left[R_{q}^{c}=1\right]-\Pr\left[R_{q}^{c'}=1\right]\right|.
\end{equation}

\end{thm}
The notion of a maximal coupling and the existence part of the theorem
above can be generalized to arbitrary systems \cite{KDL2015,conversations,DKL2015LNCS},
but in this paper we focus on the cyclic-4 systems only.

It is easy to see that if a cyclic-4 system is consistently connected,
i.e., if $\left\langle R_{q}^{c}\right\rangle =\left\langle R_{q}^{c'}\right\rangle $
for all $q,c,c'$, then in a maximal coupling $\left\{ T_{q}^{c},T_{q}^{c'}\right\} $
of any connection $\left\{ R_{q}^{c},R_{q}^{c'}\right\} $ we have
$\left\langle T_{q}^{c}T_{q}^{c'}\right\rangle =1$, or equivalently,
\begin{equation}
\Pr\left[T_{q}^{c}=T_{q}^{c'}\right]=1.
\end{equation}
In other words, in a maximal coupling the measurements in a connection
are modeled as being essentially ``the same''. This simple observation
allows us to make use of the notion of maximal couplings in the following
rigorous version of the traditional understanding of contextuality.
\begin{defn}
\label{def consistently-connected}A consistently connected cyclic-4
system (\ref{eq: matrix EPRB}) is noncontextual if it has a coupling
(\ref{eq: coupling for EPR-B}) in which $\left(S_{1}^{1},S_{1}^{4}\right),\left(S_{2}^{1},S_{2}^{2}\right),\left(S_{3}^{2},S_{3}^{3}\right),\left(S_{4}^{3},S_{4}^{4}\right)$
are maximal couplings for the corresponding connections $\left(R_{1}^{1},R_{1}^{4}\right),\left(R_{2}^{1},R_{2}^{2}\right),\left(R_{3}^{2},R_{3}^{3}\right),\left(R_{4}^{3},R_{4}^{4}\right)$,
i.e., if
\begin{equation}
\begin{array}{cc}
\Pr\left[S_{1}^{1}=S_{1}^{4}\right]=1, & \Pr\left[S_{2}^{1}=S_{2}^{2}\right]=1,\\
\Pr\left[S_{3}^{2}=S_{3}^{3}\right]=1, & \Pr\left[S_{4}^{3}=S_{4}^{4}\right]=1.
\end{array}\label{eq: identity}
\end{equation}
If such a coupling does not exist, the system is contextual.
\end{defn}
This definition allows one to preserve the spirit of the traditional
understanding (the context-irrelevance hypothesis: $R_{q}^{c}$ and
$R_{q}^{c'}$ are always ``the same'') while adhering to the logic
of the CbD approach: $R_{q}^{c}$ and $R_{q}^{c'}$ are not only different,
they are not even stochastically interrelated. From this point of
view, the following theorem, first proved mutatis mutandis by Fine
in Refs~\cite{Fine1982,Fine_PRL1982}, summarizes the traditional
analysis of contextuality for the cyclic-4 systems.
\begin{thm}
\label{thm: Fine1982}A consistently connected cyclic-4 system (\ref{eq: matrix EPRB})
is noncontextual (by Definition \ref{def consistently-connected})
if and only if (\ref{eq: CHSH}) is satisfied.
\end{thm}
This is a special case of the theorem \ref{thm: inconsistent} below,
which in turn is a special case of a theorem proved in Ref.\ \cite{KDconjecture}
(see also Refs.\ \cite{DKL2015,DKL2015LNCS,KDL2015,conversations})
that applies to a broad class of cyclic systems, of which cyclic-4
ones are a special case.

\section{\label{sec:A-general-definition}A general definition and criterion
of contextuality in the CbD framework}

The fact that we relate Definition \ref{def consistently-connected}
to the notion of maximal couplings for connections reflects the intuition
we are guided by and suggests a natural way of generalizing contextuality
beyond consistently connected systems. 

The intuition in question can be explicated as follows. For an inconsistently
connected system, we interpret the non-coincidence of the distributions
of $R_{q}^{c}$ and $R_{q}^{c'}$ as evidence that changes in context,
$c\rightarrow c'$, ``directly'' influence the measurement of $q$.
For instance, in the Alice-Bob entanglement paradigm, if the two measurements
are time-like separated, Alice's choice of the spin axis can influence
Bob's measurement along a given axis. This is referred to as ``signaling''.
It is also possible that a Charlie who receives information from both
Alice and Bob and records both their settings and their measurement
results makes systematic errors in recording Bob's results depending
on Alice's settings. This is referred to as ``context-dependent biases''.
Whatever the cause, when we model these ``direct'' influences by
a coupling $\left\{ T_{q}^{c},T_{q}^{c'}\right\} $ of $\left\{ R_{q}^{c},R_{q}^{c'}\right\} $,
we translate the differences in distributions into differences in
values: as $c$ changes into $c'$, the value of $T_{q}^{c}$ changes
into a corresponding value of $T_{q}^{c'}$. In a maximal coupling
we do this in the maximally conservative way: the values of $T_{q}^{c}$
and $T_{q}^{c'}$remain the same as often as it is allowed by their
individual distributions (in particular, they remain always the same
if the distributions are the same). Modeling by such a coupling is
always possible if $\left\{ R_{q}^{c},R_{q}^{c'}\right\} $ is coupled
in isolation. Now, if this is also possible for all the connections
taken together, within the framework of an overall coupling of the
entire system, we can say that direct influences are sufficient to
account for the system. If, however, this is not possible, then the
maximal couplings for different connections are not mutually compatible:
we interpret this as evidence that we need more than direct influences
to account for the system. This ``more'' is what we call contextuality,
as distinct from direct influences.

The generalization of Definition \ref{def consistently-connected}
to arbitrary systems of measurement therefore is straightforward:
one can simply drop the qualification ``consistently connected''
and use the general form of Theorem \ref{thm: A-maximal-coupling}.
\begin{defn}
\label{def: inconsistent}A cyclic-4 system (\ref{eq: matrix EPRB})
is noncontextual if it has a coupling (\ref{eq: coupling for EPR-B})
in which $\left(S_{1}^{1},S_{1}^{4}\right),\left(S_{2}^{1},S_{2}^{2}\right),\left(S_{3}^{2},S_{3}^{3}\right),\left(S_{4}^{3},S_{4}^{4}\right)$
are maximal couplings for the corresponding connections $\left(R_{1}^{1},R_{1}^{4}\right),\left(R_{2}^{1},R_{2}^{2}\right),\left(R_{3}^{2},R_{3}^{3}\right),\left(R_{4}^{3},R_{4}^{4}\right)$,
i.e., if
\begin{equation}
\begin{array}{c}
\Pr\left[S_{1}^{1}=S_{1}^{4}\right]=1-\left|\Pr\left[R_{1}^{1}=1\right]-\Pr\left[R_{1}^{4}=1\right]\right|,\\
\Pr\left[S_{2}^{1}=S_{2}^{2}\right]=1-\left|\Pr\left[R_{2}^{1}=1\right]-\Pr\left[R_{2}^{2}=1\right]\right|,\\
\Pr\left[S_{3}^{2}=S_{3}^{3}\right]=1-\left|\Pr\left[R_{3}^{2}=1\right]-\Pr\left[R_{3}^{3}=1\right]\right|,\\
\Pr\left[S_{4}^{3}=S_{4}^{4}\right]=1-\left|\Pr\left[R_{4}^{3}=1\right]-\Pr\left[R_{4}^{4}=1\right]\right|.
\end{array}\label{eq: general}
\end{equation}

\end{defn}
This is arguably the most conservative generalization of Definition
\ref{def consistently-connected}, but it is sufficient to deal with
all conceivable cyclic-4 systems. The correspondingly generalized
version of Theorem \ref{thm: Fine1982} is as follows \cite{DKL2015,KDconjecture,KDL2015,KujDzholdnew}.
\begin{thm}
\label{thm: inconsistent}A cyclic-4 system (\ref{eq: matrix EPRB})
is noncontextual if and only if 
\begin{equation}
\CHSH-\ICC\leq2,\label{eq:criterion}
\end{equation}
where
\begin{equation}
\CHSH=\max_{\begin{array}{c}
odd\:number\\
of-\textnormal{'s}
\end{array}}\left(\pm\left\langle R_{1}^{1}R_{2}^{1}\right\rangle \pm\left\langle R_{2}^{2}R_{3}^{2}\right\rangle \pm\left\langle R_{3}^{3}R_{4}^{3}\right\rangle \pm\left\langle R_{4}^{4}R_{1}^{4}\right\rangle \right),
\end{equation}
and
\begin{equation}
\ICC=\left|\left\langle R_{1}^{1}\right\rangle -\left\langle R_{1}^{4}\right\rangle \right|+\left|\left\langle R_{2}^{1}\right\rangle -\left\langle R_{2}^{2}\right\rangle \right|+\left|\left\langle R_{3}^{2}\right\rangle -\left\langle R_{3}^{3}\right\rangle \right|+\left|\left\langle R_{4}^{3}\right\rangle -\left\langle R_{4}^{4}\right\rangle \right|.
\end{equation}

\end{thm}
The abbreviations in this theorem are as follows. $\CHSH$ is the
left-hand side expression in the classical CHSH inequality (\ref{eq: CHSH}),
named so after the authors of Ref.\ \cite{9CHSH}. $\ICC$ is a measure
of \emph{inconsistency of the connectedness} \cite{DKL2015,KDL2015,KujDzholdnew}:
if it is zero, then the criterion (\ref{eq:criterion}) reduces to
the CHSH inequality (\ref{eq: CHSH}), and the theorem above reduces
to Theorem \ref{thm: Fine1982}.

To illustrate the computations, consider the modification of the example
of Table \ref{tab: Example 1corr} in Table \ref{tab: Example 2}.
The value of $\CHSH$ in the system is 4, the same maximal possible
value as in Table \ref{tab: Example 1corr}. But
\[
\ICC=\left|\left\langle R_{3}^{2}\right\rangle -\left\langle R_{3}^{3}\right\rangle \right|+\left|\left\langle R_{4}^{3}\right\rangle -\left\langle R_{4}^{4}\right\rangle \right|=\left|4p-2\right|,
\]
whence
\[
\CHSH-\ICC=4-\left|4p-2\right|=\left\{ \begin{array}{cc}
4\left(1-p\right)+2 & \textnormal{if }p\geq\nicefrac{1}{2},\\
4p+2 & \textnormal{if }p<\nicefrac{1}{2}.
\end{array}\right.
\]
The system is noncontextual by the criterion (\ref{eq:criterion})
only if $p=0$ or $p=1$; for other values the difference exceeds
2. 

\begin{table}
\caption{\label{tab: Example 2}An inconsistently connected cyclic-4 system.
In accordance with the general Definition \ref{def: inconsistent},
the system is contextual if and only if $p$ is not 0 or 1. In accordance
with the narrow (traditional) Definition \ref{def consistently-connected},
the notion of contextuality is not applicable unless $p=\nicefrac{1}{2}$;
in all other cases the criterion (\ref{eq: CHSH}) is not derivable
and the value of $\CHSH$ is not interpretable. \medskip{}
}

\begin{footnotesize}

\begin{centering}
\begin{tabular}{ccc|c|c|c}
 &  & \multicolumn{1}{c}{} & \multicolumn{2}{c}{$R_{2}^{1}$} & \tabularnewline
 &  & \multicolumn{1}{c}{} & \multicolumn{1}{c}{} & \multicolumn{1}{c}{} & \tabularnewline
\cline{4-5} 
 &  &  & $+1$ & $-1$ & \tabularnewline
\cline{3-6} 
\multirow{2}{*}{$R_{1}^{1}$} & \multicolumn{1}{c|}{\multirow{2}{*}{}} & $+1$ & $\nicefrac{1}{2}$ & 0 & \multicolumn{1}{c|}{$\nicefrac{1}{2}$}\tabularnewline
\cline{3-6} 
 &  & $-1$ & 0 & $\nicefrac{1}{2}$ & \multicolumn{1}{c|}{$\nicefrac{1}{2}$}\tabularnewline
\cline{3-6} 
 &  &  & $\nicefrac{1}{2}$ & $\nicefrac{1}{2}$ & \tabularnewline
\cline{4-5} 
\end{tabular}$\qquad$$\qquad$%
\begin{tabular}{c|c|c|ccc}
\multicolumn{1}{c}{} & \multicolumn{2}{c}{$R_{4}^{4}$} &  &  & \tabularnewline
\multicolumn{1}{c}{} & \multicolumn{1}{c}{} & \multicolumn{1}{c}{} &  &  & \tabularnewline
\cline{2-3} 
 & $+1$ & $-1$ &  &  & \tabularnewline
\cline{1-4} 
\multicolumn{1}{|c|}{$+1$} & $\nicefrac{1}{2}$ & 0 & \multicolumn{1}{c|}{$\nicefrac{1}{2}$} &  & \multirow{2}{*}{$R_{1}^{4}$}\tabularnewline
\cline{1-4} 
\multicolumn{1}{|c|}{$-1$} & 0 & $\nicefrac{1}{2}$ & \multicolumn{1}{c|}{$\nicefrac{1}{2}$} &  & \tabularnewline
\cline{1-4} 
 & $\nicefrac{1}{2}$ & $\nicefrac{1}{2}$ &  &  & \tabularnewline
\cline{2-3} 
\end{tabular}
\par\end{centering}

\begin{centering}
\begin{tabular}{ccc|c|c|c}
\cline{4-5} 
 &  &  & $+1$ & $-1$ & \tabularnewline
\cline{3-6} 
\multirow{2}{*}{$R_{3}^{2}$} & \multicolumn{1}{c|}{} & $+1$ & $\nicefrac{1}{2}$ & 0 & \multicolumn{1}{c|}{$\nicefrac{1}{2}$}\tabularnewline
\cline{3-6} 
 & \multicolumn{1}{c|}{} & $-1$ & 0 & $\nicefrac{1}{2}$ & \multicolumn{1}{c|}{$\nicefrac{1}{2}$}\tabularnewline
\cline{3-6} 
 &  &  & $\nicefrac{1}{2}$ & $\nicefrac{1}{2}$ & \tabularnewline
\cline{4-5} 
 &  & \multicolumn{1}{c}{} & \multicolumn{2}{c}{} & \tabularnewline
 &  & \multicolumn{1}{c}{} & \multicolumn{2}{c}{$R_{2}^{2}$} & \tabularnewline
\end{tabular}$\qquad$$\qquad$%
\begin{tabular}{c|c|c|ccc}
\cline{2-3} 
 & $+1$ & $-1$ &  &  & \tabularnewline
\cline{1-4} 
\multicolumn{1}{|c|}{$+1$} & 0 & $p$ & \multicolumn{1}{c|}{$p$} &  & \multirow{2}{*}{$R_{3}^{3}$}\tabularnewline
\cline{1-4} 
\multicolumn{1}{|c|}{$-1$} & $1-p$ & 0 & \multicolumn{1}{c|}{$1-p$} &  & \tabularnewline
\cline{1-4} 
 & $1-p$ & $p$ &  &  & \tabularnewline
\cline{2-3} 
\multicolumn{1}{c}{} & \multicolumn{2}{c}{} &  &  & \tabularnewline
\multicolumn{1}{c}{} & \multicolumn{2}{c}{$R_{4}^{3}$} &  &  & \tabularnewline
\end{tabular}
\par\end{centering}

\end{footnotesize}
\end{table}

One can see in (\ref{eq:criterion}) an algebraic realization of the
intuition described above, of direct influences being or not being
sufficient to account for the system. The direct influences are represented
by the term $\ICC$ while $\CHSH-2$ can be viewed as the total of
the dependence of measurements on contexts. If $\ICC$ is not large
enough, it does not exceed $\CHSH-2$, and in this sense it is ``insufficient''
to explain the total of the context-dependence. The difference is
the ``unexplained'' context-dependence that we view as true contextuality. 

For the arguments in favor of generalizing the definition of contextuality
to inconsistently connected systems, see Refs. \cite{conversations,KDL2015,DKL2015}.
Let us emphasize here a pragmatic argument. Since one cannot prove
a null hypothesis, dealing with experimental results one can never
be certain that consistent connectedness holds. If one confines one's
definition of contextuality to the latter case (Definition \ref{def consistently-connected}),
one's determination that a system is contextual would always be ``suspended''
and could be easily invalidated if with a larger sample size a small
inconsistency were detected. Moreover, small inconsistencies should
be expected in virtually all real experiments, as one can never be
rid of all systematic sources of error or make them perfectly counterbalanced.
None of this poses a problem for Definition \ref{def: inconsistent}:
small values of $\ICC$, unless $\CHSH$ is very close to 2, will
not change one's determination that a system is or is not contextual.

\section{\label{sec:The-Alice-Bob-version}The ``Alice-Bob'' EPR-B version
of the cyclic-4 system}

The contextuality analysis of a cyclic-4 system does not depend on
what precisely the properties $\left\{ q_{1},q_{2},q_{3},q_{4}\right\} $
are, nor on what the contexts $\left\{ c_{1},c_{2},c_{3},c_{4}\right\} $
are. All that matters is that each context involves two properties
measured ``together'', no two contexts share more than one property,
each property is measured in precisely two different contexts, and
each measurement has two possible values. 

The importance of the cyclic-4 systems, however, is primarily related
to the entanglement paradigm in quantum mechanics: two particles created
in a singular state move away from each other, reaching simultaneously
two observers, one of them Alice and another Bob; Alice chooses one
of two fixed axes and measures her particle's spin along it; Bob does
the same with his particle. Assuming the two particles are spin-$\nicefrac{1}{2}$
ones, the outcomes are binary random variables. Alice's two fixed
axes can be denoted $a_{1}=q_{1}$ and $a_{2}=q_{3}$, and Bob's axes
can be denoted $b_{1}=q_{2}$ and $b_{2}=q_{4}$. The contexts then
can be identified by the pairs of axes simultaneously chosen by Alice
and Bob: 
\begin{equation}
\begin{array}{cc}
c_{1}=\left\{ q_{1},q_{2}\right\} =\left\{ a_{1},b_{1}\right\} , & c_{2}=\left\{ q_{2},q_{3}\right\} =\left\{ b_{1},a_{2}\right\} ,\\
c_{3}=\left\{ q_{3},q_{4}\right\} =\left\{ a_{2},b_{2}\right\} , & c_{4}=\left\{ q_{4},q_{1}\right\} =\left\{ b_{2},a_{1}\right\} .
\end{array}\label{eq: A-B contexts}
\end{equation}

We can now simplify notation for the measurements by denoting Alice's
measurements by $A$ and Bob's by $B$. We will use two subscripts
of which (note the asymmetry) the first one refers to Alice's choice
of one of her two axes, and the second one refers to Bob's choice
of one of his two axes. This notation ensures that $A_{ij}$ and $B_{ij}$
(and only these, identically subscripted pairs) are jointly distributed.
Random variable $A_{ij}$ is interpreted as the outcome of measuring
property $a_{i}$ in the context of being measured together with property
$b_{j}$ (whether or not the distribution of $A_{ij}$ depends on
$j$); $B_{ij}$ is the outcome of measuring property $b_{j}$ in
the context of being measured together with property $a_{i}$ (whether
or not the distribution of $B_{ij}$ depends on $i$).

The correspondence between the general $R_{q}^{c}$ notation and the
special $A_{ij}\textnormal{-}B_{ij}$ notation is as follows: 
\begin{equation}
\begin{array}{cc}
R_{1}^{1}=R_{a_{1}}^{\left\{ a_{1},b_{1}\right\} }=A_{11}, & R_{2}^{1}=R_{b_{1}}^{\left\{ a_{1},b_{1}\right\} }=B_{11},\\
R_{2}^{2}=R_{b_{1}}^{\left\{ a_{2},b_{1}\right\} }=B_{21}, & R_{3}^{2}=R_{a_{2}}^{\left\{ a_{2},b_{1}\right\} }=A_{21},\\
R_{3}^{3}=R_{a_{2}}^{\left\{ a_{2},b_{2}\right\} }=A_{22}, & R_{4}^{3}=R_{b_{2}}^{\left\{ a_{2},b_{2}\right\} }=B_{22},\\
R_{4}^{4}=R_{b_{2}}^{\left\{ a_{1},b_{2}\right\} }=B_{12}, & R_{1}^{4}=R_{a_{1}}^{\left\{ a_{1},b_{2}\right\} }=A_{12}.
\end{array}\label{eq: correspondences}
\end{equation}

The entanglement paradigm serves as a template for other applications,
with very different meanings of the properties $a,b$ (see Ref.\ \cite{DzhRuKuj}
for examples in psychology). In this paper we will use as an example
the experiment by Aerts and Sozzo \cite{AertsSozzo2015}, where $a$
and $b$ are cardinal and intercardinal orientations chosen in the
Rose of the Winds, and the measurements are choices (by human respondents)
of one of two possible wind directions along each of these orientations,
as shown in Table \ref{tab: A=000026S}.

\begin{table}
\caption{\label{tab: A=000026S}The properties and their measurements in the
EPR-B-like system used in Ref.\ \cite{AertsSozzo2015}. In each trial
a human respondent is asked to choose ``wind directions'' along
a pair of spatial orientations in the Rose of the Winds. Each pair
consists of one cardinal orientation ($a_{1}$ or $a_{2}$) and one
intercardinal orientation ($b_{1}$ or $b_{2}$). For instance, the
respondent can be given $\left(a_{2},b_{1}\right)$, in which case
her possible choices would be ``East and Northeast'', ``East and
Southwest'', ``West and Northeast'', and ``West and Southwest'',
corresponding to four possible values of $\left(A_{21},B_{21}\right)$.
\medskip{}
}

\begin{footnotesize}

\begin{centering}
\begin{tabular}{|c|c|c|}
\hline 
$\begin{array}{c}
\textnormal{Properties, }q\\
\textnormal{(Rose of the Winds)}
\end{array}$ & $\begin{array}{c}
\textnormal{Measurements}\\
\textnormal{(Choices of direction)}
\end{array}$ & $\begin{array}{c}
\textnormal{Context}\\
\textnormal{(Options for other direction)}
\end{array}$\tabularnewline
\hline 
\hline 
North-South $\left(a_{1}\right)$ & $\begin{array}{cc}
A_{1j}=+1= & \textnormal{North}\\
A_{1j}=-1= & \textnormal{South}
\end{array}$ & \multirow{2}{*}{$\begin{array}{c}
\textnormal{measured together with }b_{j}\\
\left(j=1,2\right)
\end{array}$}\tabularnewline
\cline{1-2} 
East-West $\left(a_{2}\right)$ & $\begin{array}{cc}
A_{2j}=+1= & \textnormal{East}\\
A_{2j}=-1= & \textnormal{West}
\end{array}$ & \tabularnewline
\hline 
Northeast-Southwest $\left(b_{1}\right)$ & $\begin{array}{cc}
B_{i1}=+1= & \textnormal{Northeast}\\
B_{i1}=-1= & \textnormal{Southwest}
\end{array}$ & \multirow{2}{*}{$\begin{array}{c}
\textnormal{measured together with }a_{i}\\
\left(i=1,2\right)
\end{array}$}\tabularnewline
\cline{1-2} 
Northwest-Southeast $\left(b_{2}\right)$ & $\begin{array}{cc}
B_{i2}=+1= & \textnormal{Southeast}\\
B_{i2}=-1= & \textnormal{Northwest}
\end{array}$ & \tabularnewline
\hline 
\end{tabular}
\par\end{centering}

\end{footnotesize}
\end{table}

The results of this experiment (see Table \ref{tab: A=000026Sresults})
yield the following computations:
\[
\CHSH=2.47,\ICC=0.71,
\]
whence
\[
\CHSH-\ICC=1.76<2.
\]
We conclude that the data exhibit no contextuality in the sense of
Definition \ref{def: inconsistent}.

\begin{table}
\caption{\label{tab: A=000026Sresults}The results of the experiment reported
in Ref.\ \cite{AertsSozzo2015}. All probability estimates are computed
from polling $85$ people, treating them as $85$ realizations of
one and the same pair of random variables in each of the four contexts.
\medskip{}
}

\begin{footnotesize}

\begin{centering}
\begin{tabular}{ccc|c|c|c}
 &  & \multicolumn{1}{c}{} & \multicolumn{2}{c}{$B_{11}$} & \tabularnewline
 &  & \multicolumn{1}{c}{} & \multicolumn{1}{c}{} & \multicolumn{1}{c}{} & \tabularnewline
\cline{4-5} 
 &  &  & $+1$ & $-1$ & \tabularnewline
\cline{3-6} 
\multirow{2}{*}{$A_{11}$} & \multicolumn{1}{c|}{\multirow{2}{*}{}} & $+1$ & $0.13$ & $0.55$ & \multicolumn{1}{c|}{$0.68$}\tabularnewline
\cline{3-6} 
 &  & $-1$ & $0.25$ & $0.07$ & \multicolumn{1}{c|}{$0.32$}\tabularnewline
\cline{3-6} 
 &  &  & $0.38$ & $0.62$ & \tabularnewline
\cline{4-5} 
\end{tabular}$\qquad$$\qquad$%
\begin{tabular}{c|c|c|ccc}
\multicolumn{1}{c}{} & \multicolumn{2}{c}{$B_{12}$} &  &  & \tabularnewline
\multicolumn{1}{c}{} & \multicolumn{1}{c}{} & \multicolumn{1}{c}{} &  &  & \tabularnewline
\cline{2-3} 
 & $+1$ & $-1$ &  &  & \tabularnewline
\cline{1-4} 
\multicolumn{1}{|c|}{$+1$} & $0.47$ & $0.12$ & \multicolumn{1}{c|}{$0.59$} &  & \multirow{2}{*}{$A_{12}$}\tabularnewline
\cline{1-4} 
\multicolumn{1}{|c|}{$-1$} & $0.06$ & $0.35$ & \multicolumn{1}{c|}{$0.41$} &  & \tabularnewline
\cline{1-4} 
 & $0.53$ & $0.47$ &  &  & \tabularnewline
\cline{2-3} 
\end{tabular}
\par\end{centering}

\begin{centering}
\begin{tabular}{ccc|c|c|c}
\cline{4-5} 
 &  &  & $+1$ & $-1$ & \tabularnewline
\cline{3-6} 
\multirow{2}{*}{$A_{21}$} & \multicolumn{1}{c|}{} & $+1$ & $0.13$ & $0.38$ & \multicolumn{1}{c|}{$0.51$}\tabularnewline
\cline{3-6} 
 & \multicolumn{1}{c|}{} & $-1$ & $0.42$ & $0.07$ & \multicolumn{1}{c|}{$0.49$}\tabularnewline
\cline{3-6} 
 &  &  & $0.55$ & $0.45$ & \tabularnewline
\cline{4-5} 
 &  & \multicolumn{1}{c}{} & \multicolumn{2}{c}{} & \tabularnewline
 &  & \multicolumn{1}{c}{} & \multicolumn{2}{c}{$B_{21}$} & \tabularnewline
\end{tabular}$\qquad$$\qquad$%
\begin{tabular}{c|c|c|ccc}
\cline{2-3} 
 & $+1$ & $-1$ &  &  & \tabularnewline
\cline{1-4} 
\multicolumn{1}{|c|}{$+1$} & $0.09$ & $0.44$ & \multicolumn{1}{c|}{$0.53$} &  & \multirow{2}{*}{$A_{22}$}\tabularnewline
\cline{1-4} 
\multicolumn{1}{|c|}{$-1$} & $0.38$ & $0.09$ & \multicolumn{1}{c|}{$0.47$} &  & \tabularnewline
\cline{1-4} 
 & $0.47$ & $0.53$ &  &  & \tabularnewline
\cline{2-3} 
\multicolumn{1}{c}{} & \multicolumn{2}{c}{} &  &  & \tabularnewline
\multicolumn{1}{c}{} & \multicolumn{2}{c}{$B_{22}$} &  &  & \tabularnewline
\end{tabular}
\par\end{centering}

\end{footnotesize}
\end{table}

\section{Methodological Remarks}

One may, of course, reject the generalized Definition \ref{def: inconsistent}
and stick with the traditional understanding (Definition \ref{def consistently-connected}),
but the latter applies only to consistently connected systems of measurements,
whereas the inconsistency of the connectedness in the data of Ref.\ \cite{AertsSozzo2015}
is clearly present in spite of the small sample size used ($p<0.03$
for the difference between $\left\langle B_{11}\right\rangle $ and
$\left\langle B_{21}\right\rangle $). As explained in Section \ref{sec:Consistent-connectedness-and-contextuality},
in this case no inequality can be derived for $\CHSH$ (except for
the trivial $\CHSH\leq4$), and no interpretation is known for whether
$\CHSH$ exceeds or does not exceed any value below $4$.

The authors of Ref.\ \cite{AertsSozzo2015} are aware of the difficulties
caused by inconsistent connectedness in judging violations of the
CHSH inequality (see Ref.\ \cite{DK_Topics}), so they propose a
computational modification of their data that makes all marginal distributions
uniform. They justify this procedure by an isotropy argument, according
to which any direction in the Rose of the Winds plane could be taken
to play the role of the vector North, with all other directions rotated
to preserve their angles with respect to this new North. Using this
argument, Aerts and Sozzo average the observed probabilities in such
a way that all marginal probabilities become $\nicefrac{1}{2}$ while
the value of $\CHSH$ does not change.

An isotropy argument, however, as any other symmetry argument, only
makes sense if formulated as invariance of a relevant feature (in
our case, measurement) with respect to certain changes. To give a
trivial example, the length of a segment in the Euclidean plane is
invariant with respect to its rotations. Therefore one can average
the length measurements of a radius at different orientations, and
this averaging would only improve statistical reliability of the measurements
rather than change the true measured value. By contrast, we see in
Table \ref{tab: A=000026Sresults} that the measurements $A_{ij}$
and $B_{ij}$ are not invariant with respect to rotations: e.g., $\Pr\left[A_{11}=-1,B_{11}=1\right]$
is different from $\Pr\left[B_{21}=-1,A_{21}=1\right]$, although
the ordered pair of the orientations in the second case, $\left(b_{1},a_{2}\right)$,
is a rotated by $\pi/4$ copy of the orientation pair in the first
case, $\left(a_{1},b_{1}\right)$. Even more obvious: $\Pr\left[A_{11}=1,B_{11}=1\right]$
is not the same as $\Pr\left[A_{11}=-1,B_{11}=-1\right]$ although
they pertain to orientation pairs rotated by $\pi$ with respect to
each other.

The latter example is important for the computational modification
of the data used in Ref.\ \cite{AertsSozzo2015}. This procedure
achieves uniform marginals while retaining the value of $\CHSH$ precisely
because it considers the \emph{jointly-opposite} outcomes 
\begin{equation}
\left(A_{ij}=x,B_{ij}=y\right)\textnormal{ and }\left(A_{ij}=-x,B_{ij}=-y\right),
\end{equation}
with $x,y\in\left\{ -1,+1\right\} $, to be ``equivalent''. The
probability of each of them is therefore replaced with their average:
\begin{equation}
\frac{\Pr\left[A_{ij}=x,B_{ij}=y\right]+\Pr\left[A_{ij}=-x,B_{ij}=-y\right]}{2}.\label{eq:averaging}
\end{equation}
Let us denote by $A_{ij}^{*}$ and $B_{ij}^{*}$ the new random variables
with these symmetrized distributions. Due to the symmetry, 
\begin{equation}
\begin{array}{c}
\left\langle A_{ij}^{*}\right\rangle =\Pr\left[A_{ij}^{*}=+1\right]-\Pr\left[A_{ij}^{*}=-1\right]=0,\\
\left\langle B_{ij}^{*}\right\rangle =\Pr\left[B_{ij}^{*}=+1\right]-\Pr\left[B_{ij}^{*}=-1\right]=0.
\end{array}
\end{equation}
At the same time,
\begin{equation}
\left\langle A_{ij}^{*}B_{ij}^{*}\right\rangle =2\Pr\left[A_{ij}^{*}=B_{ij}^{*}\right]-1,
\end{equation}
and since it follows from (\ref{eq:averaging}) that
\begin{equation}
\Pr\left[A_{ij}^{*}=B_{ij}^{*}\right]=\Pr\left[A_{ij}=B_{ij}\right],
\end{equation}
the value of $\CHSH$ remains intact.

This averaging procedure has been described in the quantum physics
literature by Masanes, Acin, and Gisin \cite{masanes}; it is the
first part of their ``depolarization'' procedure. There, however,
it is meant to be a data generation or data doctoring procedure (not
a data analysis one), involving either direct signaling between Alice
or Bob, or a third party, Charley, who receives from Alice and from
Bob their settings and their measurement results, flips a fair coin,
and multiplies these measurement results (always both of them) by
$+1$ or $-1$ accordingly. Since this averaging procedure is universal
(applicable to all EPR-B systems without exception), if taken as a
data analysis procedure it amounts to ignoring the marginal probabilities
altogether and simply \emph{defining} contextuality (or entanglement)
as any violation of the CHSH inequality. 

One might ask: why not adopt this approach? It is definitely simpler
than the approach advocated by us, which involves (a) labeling the
measurements contextually, (b) determining subsystems of measurements
that are stochastically unrelated to each other, (c) defining contextuality
in terms of the (non)existence of a coupling for these subsystems
with certain constraints imposed on the connections (measurements
of the same properties in different contexts); and (d) deriving CHSH
inequalities or their generalizations as theorems \cite{conversations,DK2014Scripta,DK_PLOS_2014,DK_qualified,DKL2015,DKL2015LNCS,KDL2015,KDconjecture}.

The answer to the question is that adopting the definition in question,
in addition to being arbitrary, would make construction of contextual
systems child's play: the contextual system will become ubiquitous
and obvious, including systems in classical mechanics and human behavior
that no one normally would think of as contextual. Moreover, with
the definition in question one would have to forget about the ``quantum''
motivation for seeking contextuality, because these contextual systems
in classical mechanics and human behavior would violate Tsirelson
(or Cirel'son) bounds \cite{landau,Cirel'son} as easily as they would
the CHSH ones.

\section{Contextuality as child's play}

We will consider just one example, with multiple possible implementations.
Table \ref{tab: child's play} represents the probabilities of $\left[A_{ij}=x,B_{ij}=y\right]$
in a hypothetical EPR-B-type system ($i,j\in\left\{ 1,2\right\} $,
$x,y\in\left\{ -1,+1\right\} $). Here, 
\[
\CHSH=4,
\]
the algebraically maximal possible value for $\CHSH$. The system
is, however, noncontextual by Definition \ref{def: inconsistent}
and Theorem~\ref{thm: inconsistent}: $\ICC$ in it equals the value
of $\left\langle A_{21}\right\rangle -\left\langle A_{22}\right\rangle =2$,
whence
\[
\CHSH-\ICC=2
\]
In fact, any deterministic system (one in which all probabilities
are 0 or 1) is noncontextual. A simple way of demonstrating this is
as follows: a deterministic system has a single coupling, and its
subcouplings corresponding to connections (each of which is deterministic)
are their only couplings, hence maximal ones.

However, if in Table \ref{tab: child's play} one decides to ignore
marginal probabilities, the system is maximally contextual (and in
fact more contextual than allowed by quantum mechanics).

\begin{table}
\caption{\label{tab: child's play}A deterministic system that is (as any other
deterministic system) noncontextual by Definition \ref{def: inconsistent}
but is ``maximally contextual'' if one ignores marginal probabilities
(or, equivalently, averages over jointly-opposite outcomes). \medskip{}
}

\begin{footnotesize}

\begin{centering}
\begin{tabular}{ccc|c|c|c}
 &  & \multicolumn{1}{c}{} & \multicolumn{2}{c}{$B_{11}$} & \tabularnewline
 &  & \multicolumn{1}{c}{} & \multicolumn{1}{c}{} & \multicolumn{1}{c}{} & \tabularnewline
\cline{4-5} 
 &  &  & $+1$ & $-1$ & \tabularnewline
\cline{3-6} 
\multirow{2}{*}{$A_{11}$} & \multicolumn{1}{c|}{\multirow{2}{*}{}} & $+1$ & $1$ & $0$ & \multicolumn{1}{c|}{$1$}\tabularnewline
\cline{3-6} 
 &  & $-1$ & $0$ & $0$ & \multicolumn{1}{c|}{$0$}\tabularnewline
\cline{3-6} 
 &  &  & $1$ & $0$ & \tabularnewline
\cline{4-5} 
\end{tabular}$\qquad$$\qquad$%
\begin{tabular}{c|c|c|ccc}
\multicolumn{1}{c}{} & \multicolumn{2}{c}{$B_{12}$} &  &  & \tabularnewline
\multicolumn{1}{c}{} & \multicolumn{1}{c}{} & \multicolumn{1}{c}{} &  &  & \tabularnewline
\cline{2-3} 
 & $+1$ & $-1$ &  &  & \tabularnewline
\cline{1-4} 
\multicolumn{1}{|c|}{$+1$} & $1$ & $0$ & \multicolumn{1}{c|}{$1$} &  & \multirow{2}{*}{$A_{12}$}\tabularnewline
\cline{1-4} 
\multicolumn{1}{|c|}{$-1$} & $0$ & $0$ & \multicolumn{1}{c|}{$0$} &  & \tabularnewline
\cline{1-4} 
 & $1$ & $0$ &  &  & \tabularnewline
\cline{2-3} 
\end{tabular}
\par\end{centering}

\begin{centering}
\begin{tabular}{ccc|c|c|c}
\cline{4-5} 
 &  &  & $+1$ & $-1$ & \tabularnewline
\cline{3-6} 
\multirow{2}{*}{$A_{21}$} & \multicolumn{1}{c|}{} & $+1$ & $1$ & $0$ & \multicolumn{1}{c|}{$1$}\tabularnewline
\cline{3-6} 
 & \multicolumn{1}{c|}{} & $-1$ & $0$ & $0$ & \multicolumn{1}{c|}{$0$}\tabularnewline
\cline{3-6} 
 &  &  & $1$ & $0$ & \tabularnewline
\cline{4-5} 
 &  & \multicolumn{1}{c}{} & \multicolumn{2}{c}{} & \tabularnewline
 &  & \multicolumn{1}{c}{} & \multicolumn{2}{c}{$B_{21}$} & \tabularnewline
\end{tabular}$\qquad$$\qquad$%
\begin{tabular}{c|c|c|ccc}
\cline{2-3} 
 & $+1$ & $-1$ &  &  & \tabularnewline
\cline{1-4} 
\multicolumn{1}{|c|}{$+1$} & $0$ & $0$ & \multicolumn{1}{c|}{0} &  & \multirow{2}{*}{$A_{22}$}\tabularnewline
\cline{1-4} 
\multicolumn{1}{|c|}{$-1$} & $1$ & $0$ & \multicolumn{1}{c|}{$1$} &  & \tabularnewline
\cline{1-4} 
 & $1$ & $0$ &  &  & \tabularnewline
\cline{2-3} 
\multicolumn{1}{c}{} & \multicolumn{2}{c}{} &  &  & \tabularnewline
\multicolumn{1}{c}{} & \multicolumn{2}{c}{$B_{22}$} &  &  & \tabularnewline
\end{tabular}
\par\end{centering}

\end{footnotesize}
\end{table}

It is trivial to find or construct a system described by Table \ref{tab: child's play}.
To begin with conceptual combinations, consider, e.g., the experiment
in which the properties $a,b$ and measurements $A,B$ are identified
as shown in Table \ref{tab: child's play 1}. Such an experiment would
yield Table \ref{tab: child's play} unless the participants choose
to deliberately give wrong responses. There is, in fact, nothing wrong
in considering the conceptual inferences like ``Green Triangle is
Green'' and ``Green Triangle is Triangular'' as examples of contextuality
or ``(super-quantum) entanglement,'' but this looks to us as making
the concept of contextuality too trivial to be of interest. 

\begin{table}
\caption{\label{tab: child's play 1}A trivial system described by Table \ref{tab: child's play}
if people are instructed to choose correct responses. In the context
$\left(a_{1},b_{1}\right)$ the choice is among the sentences: ``Green
Triangle is Green'', ``Green Triangle is Red'', ``Yellow Circle
is Green'', and ``Yellow Circle is Red''. In the context $\left(a_{1},b_{2}\right)$
the choice is among the sentences: ``Green Triangle is Triangular'',
``Green Triangle is Square'', ``Yellow Circle is Triangular'',
and ``Yellow Circle is Square''. In the context $\left(a_{2},b_{1}\right)$
the choice is among the sentences: ``Green Circle is Green'', ``Green
Circle is Red'', ``Yellow Triangle is Green'', and ``Yellow Triangle
is Red''. In the context $\left(a_{2},b_{2}\right)$, the choice
is among the sentences: ``Green Circle is Triangular'', ``Green
Circle is Square'', ``Yellow Triangle is Triangular'', and ``Yellow
Triangle is Square''. \medskip{}
}

\begin{footnotesize}

\begin{centering}
\begin{tabular}{|c|c|c|}
\hline 
$\begin{array}{c}
\textnormal{Properties}\\
\textnormal{(choices between)}
\end{array}$ & $\begin{array}{c}
\textnormal{Measurements}\\
\textnormal{(choices)}
\end{array}$ & $\textnormal{Context}$\tabularnewline
\hline 
\hline 
Green Triangle and Yellow Circle $\left(a_{1}\right)$ & $\begin{array}{c}
A_{1j}=+1=\textnormal{\textnormal{Green Triangle}}\\
A_{1j}=-1=\textnormal{Yellow Circle}
\end{array}$ & \multirow{2}{*}{$\begin{array}{c}
\textnormal{together with }b_{j}\\
\left(j=1,2\right)
\end{array}$}\tabularnewline
\cline{1-2} 
Green Circle and Yellow Triangle $\left(a_{2}\right)$ & $\begin{array}{c}
A_{2j}=+1=\textnormal{Green Circle}\\
A_{2j}=-1=\textnormal{Yellow Triangle}
\end{array}$ & \tabularnewline
\hline 
Green and Red $\left(b_{1}\right)$ & $\begin{array}{c}
B_{i1}=+1=\textnormal{Green}\\
B_{i1}=-1=\textnormal{Red}
\end{array}$ & \multirow{2}{*}{$\begin{array}{c}
\textnormal{together with }a_{i}\\
\left(j=1,2\right)
\end{array}$}\tabularnewline
\cline{1-2} 
Triangular and Square $\left(b_{2}\right)$ & $\begin{array}{c}
B_{i2}=+1=\textnormal{Triangular}\\
B_{i2}=-1=\textnormal{Square}
\end{array}$ & \tabularnewline
\hline 
\end{tabular}
\par\end{centering}

\end{footnotesize}
\end{table}

Another example of ``conceptual entanglement'' involves creation
of new concepts in children by means of teaching them a simple nonsense
verse:

\begin{spacing}{0.5}
\begin{footnotesize}%
\begin{minipage}[t]{1\columnwidth}%
\begin{center}
\medskip{}
Pips and Nips are Zops, not Zogs.\\Pops and Nops aren't Zops nor
Zogs.\\Pips and Nops are Gots, not Gons.\\Pops and Nips aren't Gots
nor Gons.
\par\end{center}

\begin{center}
\medskip{}

\par\end{center}%
\end{minipage}\end{footnotesize}
\end{spacing}

Children who learned this piece of poetry by heart (or are allowed
to look at it while responding) would confidently respond to the questions
like ``Are Pips Zops?'' and ``Are Pops Gots?'' The resulting table
of the probabilities for them will be the same as in Table \ref{tab: child's play},
on denoting the conditions and outcomes as in Table \ref{tab: child's play-2}.

\begin{table}
\caption{\label{tab: child's play-2}A trivial system described by Table \ref{tab: child's play}
if people are instructed to choose correct responses given by the
nonsense verse about Pips, Nips, etc. \medskip{}
}

\begin{footnotesize}

\begin{centering}
\begin{tabular}{|c|c|c|}
\hline 
$\begin{array}{c}
\textnormal{Properties}\\
\textnormal{(choices between)}
\end{array}$ & $\begin{array}{c}
\textnormal{Measurements}\\
\textnormal{(choices)}
\end{array}$ & $\textnormal{Context}$\tabularnewline
\hline 
\hline 
Pip and Pop $\left(a_{1}\right)$ & $\begin{array}{cc}
A_{1j}=+1= & \textnormal{Pip}\\
A_{1j}=-1= & \textnormal{Pop}
\end{array}$ & \multirow{2}{*}{$\begin{array}{c}
\textnormal{together with }b_{j}\\
\left(j=1,2\right)
\end{array}$}\tabularnewline
\cline{1-2} 
Nip and Nop $\left(a_{2}\right)$ & $\begin{array}{cc}
A_{2j}=+1= & \textnormal{Nip}\\
A_{2j}=-1= & \textnormal{Nop}
\end{array}$ & \tabularnewline
\hline 
Zop and Zog $\left(b_{1}\right)$ & $\begin{array}{cc}
B_{i1}=+1= & \textnormal{Zop}\\
B_{i1}=-1= & \textnormal{Zog}
\end{array}$ & \multirow{2}{*}{$\begin{array}{c}
\textnormal{together with }a_{i}\\
\left(i=1,2\right)
\end{array}$}\tabularnewline
\cline{1-2} 
Got and Gon $\left(b_{2}\right)$ & $\begin{array}{cc}
B_{i2}=+1= & \textnormal{Got}\\
B_{i2}=-1= & \textnormal{Gon}
\end{array}$ & \tabularnewline
\hline 
\end{tabular}
\par\end{centering}

\end{footnotesize}
\end{table}

Finally, here is a scenario of creating Table \ref{tab: child's play}
in a purely classical physical situation. There is a gadget ``Alice''
that responds to inputs $i,j\in\left\{ 1,2\right\} $ by computing
$A=\min\left(1,2i+2j-5\right)$, and a gadget ``Bob'' that outputs
$1$ no matter what. This example is essentially identical to one
given by Filk in Figure 3 of Ref. \cite{Filk}. No physicist, as it
seems to us, would call the system consisting of these two gadgets
entangled or contextual. It is simply that both inputs influence one
of the outputs (Alice's), resulting in the observed inconsistent connectedness.

\section{Conclusion}

Inconsistent connectedness is almost a universal rule in behavioral
and social data (e.g., it is very plausible that the task of choosing
between the North and South winds affects the probabilities with which
one, in the same trial, chooses between the Northeast and Southwest
winds). It is therefore a sound scientific strategy to make inconsistent
connectedness part of one's theory of contextuality. Inconsistent
connectedness means that the measurement of a property is directly
influenced by the measurement of other properties, and this may or
may not be sufficient to account for a system's behavior. For instance,
in the experiment described in Ref.\ \cite{Lapkiewicz2011} we find
violations of consistent connectedness due to context-dependent biases
in measurements, but the detailed analysis presented in Ref.\ \cite{KDL2015}
shows that contextuality, in the sense of Definition \ref{def: inconsistent},
is still prominently present in these data. By contrast, the system
in Table \ref{tab: child's play} is noncontextual by Definition \ref{def: inconsistent},
which means that the direct influences it entails are sufficient to
explain its behavior (no contextuality exists ``on top of'' these
input-output relations). The same conclusion applies to the many different
experiments analyzed in Ref.\ \cite{DzhRuKuj} and to the experiment
reported in Ref.\ \cite{AertsSozzo2015}. We have argued that the
justification proposed in the latter for averaging across different
contexts is not tenable, and the reason it works as desired is that
it is equivalent to ignoring marginal probabilities altogether. The
consequence of such ignoring, in addition to being ad hoc, is that
contextuality becomes trivial and uninteresting. 

We make no claim, however, that contextuality, in the sense of our
definition, cannot be found in behavioral data: we merely say that
we have not found it yet. We also acknowledge that there may be viable
alternatives to our Definition \ref{def: inconsistent} that also
take into account inconsistent connectedness in a different way.

Finally, we would like to refer the reader to the concluding part
of Ref.\ \cite{DzhRuKuj} to emphasize that absence of contextuality
in behavioral and social systems \emph{does} \emph{not} mean that
quantum formalisms are not applicable to them. The so-called QQ equality,
in our opinion the most impressive outcome of quantum cognition research
to date \cite{Wang,Wang-Busemeyer}, provides a clear illustration
of how absence of contextuality can in fact be precisely a prediction
derived from quantum theory.

\section*{Acknowledgments}

This research has been supported by NSF grant SES-1155956, AFOSR grant
FA9550-14-1-0318, A. von Humboldt Foundation, and by the J. William
Fulbright Grant from Fulbright Colombia.


\begin{thebibliography}{10}
\bibitem{DKL2015}Dzhafarov EN, Kujala JV, Larsson J-Å. 2015 Contextuality
in three types of quantum-mechanical systems. \emph{Found. Phys. }\textbf{7,}\emph{
}762--782. (doi:10.1007/s10701-015-9882-9)

\bibitem{DKL2015LNCS}Dzhafarov EN, Kujala JV, Cervantes VH. 2016
Contextuality-by-Default: A brief overview of ideas, concepts, and
terminology. \emph{Lect. Notes Comput. Sc. }\textbf{9535}, 12\textendash 23.

\bibitem{DzhRuKuj}Dzhafarov EN, Zhang R, Kujala JV. 2015 Is there
contextuality in behavioral and social systems? \emph{Phil. Trans.
R. Soc. A} \textbf{374}, 20150099 (doi:10.1098/rsta.2015.0099).

\bibitem{KDconjecture}Kujala JV, Dzhafarov EN. 2015 Proof of a conjecture
on contextuality in cyclic systems with binary variables. arXiv:1503.02181.

\bibitem{KDL2015}Kujala JV, Dzhafarov EN, Larsson J-Å. 2015 Necessary
and sufficient conditions for maximal noncontextuality in a broad
class of quantum mechanical systems. \emph{Phys. Rev. Lett.} \textbf{115,}
150401. (doi:10.1103/PhysRevLett.115.150401)

\bibitem{AertsSozzo2015}Aerts D, Sozzo S. 2015 Spin and wind directions:
Identifying entanglement in nature and cognition. arXiv:1508.00434v2.

\bibitem{DK_Topics}Dzhafarov EN, Kujala, JV. 2014 Selective influences,
marginal selectivity, and Bell/CHSH inequalities. \emph{Top. Cogn.
Sci.} \textbf{6,} 121--128. (doi:10.1111/tops.12060)

\bibitem{Aerts}Aerts D, Gabora L, Sozzo S. 2013 Concepts and their
dynamics: A quantum-theoretic modeling of human thought. \emph{Top.
Cogn. Sci. }\textbf{5,} 737--772. (doi:10.1111/tops.12042)

\bibitem{DK_qualified}Dzhafarov EN, Kujala JV. 2014 A qualified Kolmogorovian
account of probabilistic contextuality. \emph{Lect. Notes Comput.
Sc.} \textbf{8369,} 201--212. (doi:10.1007/978-3-642-54943-4\_18)

\bibitem{DK_PLOS_2014}Dzhafarov EN, Kujala JV. 2014 Embedding quantum
into classical: Contextualization vs conditionalization. \emph{PLoS
One }\textbf{9, }e92818. (doi:10.1371/journal.pone.0092818)

\bibitem{DK2014Scripta}Dzhafarov EN, Kujala JV. 2014 Contextuality
is about identity of random variables.\emph{ Phys. Scripta} \textbf{T163,}
014009. (doi:10.1088/0031-8949/2014/T163/014009)

\bibitem{KujDzholdnew}Kujala JV, Dzhafarov EN. 2015. Probabilistic
Contextuality in EPR/Bohm-type systems with signaling allowed. In
\emph{Contextuality from Quantum Physics to Psychology} (eds EN Dzhafarov,
JS Jordan, R Zhang, VH Cervantes), pp. 287-308, New Jersey:World Scientific
Press.

\bibitem{PopescuRohrlich}Popescu S, Rohrlich D. 1994 Quantum nonlocality
as an axiom. \emph{Found. Phys.} \textbf{24,} 379--385. (doi:10.1007/BF02058098)

\bibitem{DK2012LNCS}Dzhafarov EN, Kujala JV. 2012 Quantum entanglement
and the issue of selective influences in psychology: An overview.
\emph{Lect. Notes Comput. Sc.} \textbf{7620,} 184--195. (doi:10.1007/978-3-642-35659-9\_17)

\bibitem{TownSchw1989}Townsend JT, Schweickert R. 1989 Toward the
trichotomy method of reaction times: Laying the foundation of stochastic
mental networks. \emph{J. Math. Psychol}. \textbf{33,} 309--327. (doi:10.1016/0022-2496(89)90013-8)

\bibitem{Ramanathan2012}Ramanathan R, Soeda A, Kurzynski P, Kasznlikowski
D. 2012 Generalized Monogamy of Contextual Inequalities from the No-Disturbance
Principle. \emph{Phys. Rev. Lett.} \textbf{109,} 050404. (doi:10.1103/PhysRevLett.109.050404)

\bibitem{Cereceda2000}Cereceda J. 2000 Quantum mechanical probabilities
and general probabilistic constraints for Einstein-Podolsky-Rosen-Bohm
experiments. \emph{Found. Phys. Lett.} \textbf{13,} 427--442. (doi:10.1023/A:1007828731477)

\bibitem{Bell1964}Bell JS. 1964 On the Einstein-Podolsky-Rosen paradox.
\emph{Physics} \textbf{1,} 195--200.

\bibitem{Bell1966}Bell JS. 1966 On the problem of hidden variables
in quantum mechanics. \emph{Rev. Mod. Phys.} \textbf{38,} 447--453.
(doi:10.1103/RevModPhys.38.447)

\bibitem{9CHSH}Clauser JF, Horne MA, Shimony A, Holt RA. 1969 Proposed
experiment to test local hidden-variable theories. \emph{Phys. Rev.
Lett.} \textbf{23,} 880--884. (doi:10.1103/PhysRevLett.23.880)

\bibitem{conversations}Dzhafarov EN, Kujala JV. 2015 Conversations
on contextuality. In \emph{Contextuality from Quantum Physics to Psychology}
(eds EN Dzhafarov, JS Jordan, R Zhang, VH Cervantes), pp. 1-22, New
Jersey: World Scientific Press.

\bibitem{Fine_PRL1982}Fine A. 1982 Hidden variables, joint probability,
and the Bell inequalities.\emph{ Phys. Rev. Lett.} \textbf{48,} 291--295.
(doi:10.1103/PhysRevLett.48.291)

\bibitem{Fine1982}Fine A. 1982 Joint distributions, quantum correlations,
and commuting observables.\emph{ J. Math. Phys.} \textbf{23,} 1306--1310.
(doi:10.1063/1.525514)

\bibitem{masanes}Masanes Ll, Acin A, Gisin N. 2006 General properties
of nonsignaling theories. \emph{Phys. Rev. A} \textbf{73,} 012112.
(doi:10.1103/PhysRevA.73.012112)

\bibitem{Cirel'son}Cirel\textquoteright son BS. 1980 Quantum generalizations
of Bell\textquoteright s inequality. \emph{Lett. Math. Phys.} \textbf{4},
93\textendash 100.

\bibitem{landau}Landau LJ. 1987 On the violation of Bell's inequality
in quantum theory. \emph{Phys. Lett. A} \textbf{120,} 54--56. (doi:10.1016/0375-9601(87)90075-2)

\bibitem{Filk}Filk T. 2015. It is the theory which decides what we
can observe. In \emph{Contextuality from Quantum Physics to Psychology}
(eds EN Dzhafarov, JS Jordan, R Zhang, VH Cervantes), pp. 77-92, New
Jersey:World Scientific Press.

\bibitem{Lapkiewicz2011}Lapkiewicz R , Li P , Schaeff C, Langford
NK, Ramelow S, Wie\'{s}niak M, Zeilinger A. 2011 Experimental non-classicality
of an indivisible quantum system. \emph{Nature} \textbf{474,} 490--493.
(doi:10.1038/nature10119)

\bibitem{Wang}Wang Z, Solloway T, Shiffrin RM, Busemeyer JR. 2014
Context effects produced by question orders reveal quantum nature
of human judgments. \emph{P. Natl. Acad. Sci. USA} \textbf{111,} 9431--9436.
(doi:10.1073/pnas.1407756111)

\bibitem{Wang-Busemeyer}Wang Z, Busemeyer JR. 2013 A quantum question
order model supported by empirical tests of an a priori and precise
prediction. \emph{Top. Cogn. Sci.} \textbf{5,} 689--710. (doi:10.1111/tops.12040)\end{thebibliography}
\end{document}